\documentclass[prl,twocolumn,longbibliography,superscriptaddress,preprintnumbers]{revtex4-2}
\usepackage{amsmath}
\usepackage{amssymb}
\usepackage{bm}
\usepackage{braket}
\usepackage{amsthm}
\usepackage{diagbox}
\usepackage{setspace}
\usepackage{verbatim}
\usepackage{esint}
\usepackage{physics}
\usepackage{indentfirst}
\usepackage{graphicx}
\usepackage{float}
\usepackage{dsfont}
\usepackage{mathrsfs}
\usepackage{mathtools}
\usepackage{tikz-cd}
\usepackage{simplewick}
\usepackage{slashed}
\usepackage{mathtools}
\usepackage{hyperref}
\usepackage{slashed}
\usepackage{cancel}
\usepackage{xcolor}
\usepackage[normalem]{ulem}
\usepackage{youngtab}
\usepackage{ytableau}
\usepackage{enumerate}
\usepackage{svg}
\usepackage[T5]{fontenc}

\let \oldbm \bm
\renewcommand{\vec}[1]{\oldbm{#1}}

\def\bk{{\vec k}}

\def\bA{{\vec A}}

\def\bs{{\vec s}}
\def\bS{{\vec S}}

\def\bq{{\vec q}}

\def\bR{{\vec R}}
\def\bG{{\vec G}}

\def\bn{{\vec n}}
\def\bm{{\vec m}}

\def\br{{\vec r}}

\def\bsigma{{\boldsymbol \sigma}}

\def\sgn{\mathop{\mathrm{sgn}}}
\def\tr{\mathop{\mathrm{tr}}}

\def\T{\mathcal{T}}

\def\D{\mathcal{D}}

\def\H{\mathcal{H}}

\def\U{{\rm U}}

\def\SU{{\rm SU}}

\newcommand{\beq}{\begin{equation}}
\newcommand{\eeq}{\end{equation}}
\newcommand{\beqarray}{\begin{eqnarray}}
\newcommand{\eeqarray}{\end{eqnarray}}

\makeatletter
\newcommand\RedeclareMathOperator{%
  \@ifstar{\def\rmo@s{m}\rmo@redeclare}{\def\rmo@s{o}\rmo@redeclare}%
}
\newcommand\rmo@redeclare[2]{%
  \begingroup \escapechar\m@ne\xdef\@gtempa{{\string#1}}\endgroup
  \expandafter\@ifundefined\@gtempa
     {\@latex@error{\noexpand#1undefined}\@ehc}%
     \relax
  \expandafter\rmo@declmathop\rmo@s{#1}{#2}}
\newcommand\rmo@declmathop[3]{%
  \DeclareRobustCommand{#2}{\qopname\newmcodes@#1{#3}}%
}
\@onlypreamble\RedeclareMathOperator
\makeatother
\RedeclareMathOperator{\Im}{Im}
\RedeclareMathOperator{\Re}{Re}

\makeatletter
\let\oldabs\abs
\def\abs{\@ifstar{\oldabs}{\oldabs*}}
\let\oldnorm\norm
\def\norm{\@ifstar{\oldnorm}{\oldnorm*}}
\makeatother

\begin{document}

\title{Untwisting moir\'e physics: Almost ideal bands and fractional Chern insulators in periodically strained monolayer graphene}
\author{Qiang Gao}
\affiliation{Department of Physics, The University of Texas at Austin, TX 78712, USA}
\author{Junkai Dong}
\affiliation{Department of Physics, Harvard University, Cambridge, MA 02138, USA}
\author{Patrick Ledwith}
\affiliation{Department of Physics, Harvard University, Cambridge, MA 02138, USA}
\author{Daniel Parker}
\affiliation{Department of Physics, Harvard University, Cambridge, MA 02138, USA}
\author{Eslam Khalaf}
\affiliation{Department of Physics, The University of Texas at Austin, TX 78712, USA}
\date{\today}
\begin{abstract}

Moir\'e systems have emerged in recent years as a rich platform to study strong correlations. Here, we will discuss a simple, experimentally feasible setup based on periodically strained graphene that reproduces several key aspects of twisted moir\'e heterostructures --- but without introducing a twist. We consider a monolayer graphene sheet subject to a $C_2$-breaking periodic strain-induced psuedomagnetic field (PMF) with period  $L_M \gg a$, along with a scalar potential of the same period. This system has {\it almost ideal} flat bands with valley-resolved Chern number $\pm 1$, where the deviation from ideal band geometry is analytically controlled and exponentially small in the dimensionless ratio $(L_M/l_B)^2$ where $l_B$ is the magnetic length corresponding to the maximum value of the PMF.  Moreover, the scalar potential can tune the bandwidth far below the Coulomb scale, making this a very promising platform for strongly interacting topological phases. Using a combination of strong-coupling theory and self-consistent Hartree fock, we find quantum anomalous Hall states at integer fillings. At fractional filling, exact diagonaliztion reveals a fractional Chern insulator at parameters in the experimentally feasible range. Overall, we find that this system has larger interaction-induced gaps, smaller quasiparticle dispersion, and enhanced tunability compared to twisted graphene systems, even in their ideal limit.

\end{abstract}
\maketitle

\emph{Introduction}--- 
The discovery of correlated states in moir\'e materials has transformed the study of strongly correlated phases~\cite{Cao_2019,Yankowitz_2019,Lu_2019,Stepanov_2020,Cao_2021,Liu_2021}. Moir\'e materials provide a platform where the bandwidth can be tuned by adjusting the twist angle, enabling the realization of topologically trivial and non-trivial strongly interacting bands. 
Beyond bandwidth and topology, recent works have identified the quantum geometry of the wavefunctions \cite{WuDasSarma, ledwithFractionalChernInsulator2020a,ledwithStrongCouplingTheory2021,khalafSoftModesMagic2020, wangExactLandauLevel2021b, abouelkomsan_quantum_2022} as a central ingredient in understanding interacting physics, including the effective quasiparticle dispersion \cite{RepellinYahui, TBGV, KangBernevigVafek,abouelkomsan_quantum_2022}, the stability of correlated topological phases \cite{ledwithFractionalChernInsulator2020a,repellin_chern_2020,abouelkomsanParticleHoleDualityEmergent2020a,wilhelmInterplayFractionalChern2021a,ledwithStrongCouplingTheory2021,parkerFieldtunedZerofieldFractional2021} and the type and properties of collective excitations \cite{WuDasSarma,khalafSoftModesMagic2020, TBGV, khalaf2021Polaron, Kwan2022skyrmions, SchindlerTrions}.
However, compared to bandwidth, quantum geometry is significantly more difficult to tune since it is mostly fixed by the form of the moir\'e potential.

A prominent example is twisted bilayer graphene (TBG), where an ideal limit called the chiral limit \cite{tarnopolsky2019origin} can be theoretically achieved by tuning intrasublattice moir\'e tunneling to zero. The model exhibits flat $C = \pm 1$ bands satisfying the trace condition \cite{ledwithFractionalChernInsulator2020a, ledwithStrongCouplingTheory2021, wangExactLandauLevel2021b, vortexability}, which relates the Fubini-study metric to the Berry curvature. Such bands, which have been dubbed ``ideal bands", are equivalent to those of the lowest Landau level (LLL) in a non-uniform magnetic field \cite{ledwithFractionalChernInsulator2020a, ledwithStrongCouplingTheory2021}, making them a very promising platform to realize\cite{xieFractionalChernInsulators2021a} exotic phases such as fractional Chern insulators (FCIs) \cite{ledwithFractionalChernInsulator2020a,repellin_chern_2020,abouelkomsanParticleHoleDualityEmergent2020a,repellin_chern_2020,wilhelmInterplayFractionalChern2021a,ledwithStrongCouplingTheory2021,parkerFieldtunedZerofieldFractional2021} and skyrmion superconductivity \cite{khalafChargedSkyrmionsTopological2021,chatterjeeSkyrmionSuperconductivityDMRG2020}.
However, known experimental knobs  cannot tune TBG to its chiral (ideal) limit (although lattice relaxation moves couplings towards this limit \cite{namLatticeRelaxationEnergy2017a, Carr2018relax, TBorNotTB}). Alternating-twist multilayer generalizations \cite{khalafMagicAngleHierarchy2019, KimTrilayer, park2021tunable, park2022robust, zhang2022promotion} may improve the situation, particularly at higher magic angles \cite{TBorNotTB}, but still do not offer sufficient tunability. Other moir\'e systems employing Bernal-stacked bilayer graphene such as twisted mono-bilayer \cite{MonobilayerYankowitz,monobi2,MonobilayerYoung,monobi4,polshynElectricalSwitchingMagnetic2020,monobiLi_STM,monobiTong_STM} or double-bilayer \cite{bibi1,bibi2,bibi3,bibi4,bibi5,bibi6,bibi7,bibi8Liu} admit idealized chiral models \cite{ledwithFamilyIdealChern2022,wangHierarchyIdealFlatbands2022a,HigherChern,wangOriginofModel} but in practice involve additional terms such as trigonal warping \cite{leeTheoryCorrelatedInsulating2019} which moves them even further from ideal conditions.

Strain engineering provides another route to realize narrow bands with strong correlations ~\cite{ghaemi2012fractional,bi2019designing,lau2021designing,yang2022origami}. 
Strain acts on graphene as a pseudo-magnetic field (PMF) with equal and opposite strength in each valley \cite{suzuura2002phonons,manesSymmetrybasedApproachElectronphonon2007,kimGrapheneElectronicMembrane2008,guineaGaugeFieldInduced2008,pereiraStrainEngineeringGraphene2009,vozmedianoGaugeFieldsGraphene2010,de2012space,manesGeneralizedEffectiveHamiltonian2013,de2013gauge}.
Early theoretical works focused on strain profiles that realize a uniform PMF to emulate Landau level physics \cite{pereiraStrainEngineeringGraphene2009,guinea2010energy,low2010strain}. However, these realizations require the atomic displacement $u$ to grow quadratically with distance~\cite{footnote1} 
which is only possible experimentally within a limited length scale ($\sim 10-100\rm{nm}$)~\cite{levy2010strain,li2020valley}.
A more controllable setup is that of periodic strain, which yields a periodic PMF with a vanishing average over the unit cell. This is realized experimentally by suspending monolayer graphene on a network of nanorods \cite{Nanorods}, or through the spontaneous buckling of a graphene sheet on specific substrates such as NbSe${}_2$ where a $C_2$-breaking PMF was recently observed~\cite{mao2020evidence}.
This PMF was shown to give rise to narrow bands \cite{phong2022boundary,de2022network, BandFlatteningBuckled, CorrelationsBuckled}, but their quantum geometry and the resulting interaction physics are yet to be explored.

Recent progress in understanding the conditions for ideal bands in Dirac systems was inspired by Ref.~\cite{tarnopolsky2019origin}, which identified general conditions for ideal flat bands in chirally symmetric Dirac Hamiltonians. A fully flat ideal band is realized if the sublattice-polarized wavefunctions at the Dirac point have zeros in real space \cite{tarnopolsky2019origin, Sheffer2022symmetries}. However, one important distinction between strain and moir\'e potentials is that the former gives rise to an {\it Abelian} gauge field whereas the latter gives a {\it non-Abelian} gauge field \cite{GuineaNonAbelian, tarnopolsky2019origin} for the Dirac electrons. This 
poses a challenge for the realization of ideal bands in strained graphene, since the sublattice polarized wavefunctions of a Dirac particle in an Abelian field are exponential functions that can never have zeros. 

In this letter, we will show that by combining slowly-varying periodic $C_2$-breaking PMF with a scalar potential of the same periodicity in monolayer graphene, we can realize an {\it almost ideal} flat band with valley resolved Chern number $C=\pm 1$. By {\it almost ideal}, we means that deviations from ideality, i.e. trace condition violation, are analytically controlled and exponentially small $\sim e^{-\alpha}$, where $\alpha \sim (L_M/l_B)^2$. Here, $L_M \gg a_{\rm graphene}$ is the period of the PMF and $l_B$ is the magnetic length corresponding to the maximal PMF. This deviation is $\ll 1$ for experimentally realistic parameters.

We show that the bandwidth is tunable by tuning the scalar field, and can be made almost two orders of magnitude smaller than the Coulomb scale. We study this limit of small bandwidth, where the interaction is expected to dominate the physics, using analytical strong coupling theory,  Hartree-Fock and exact diagonalization. We provide evidence for quantum anomalous Hall (QAH) states and fractional Chern insulators (FCIs) at integer and fractional fillings, respectively. Our results suggest that this system is more tunable and has favorable parameters to realize QAH and FCI states compared to twisted graphene systems, even in their ideal limit. 

\begin{figure}
    \centering
    \includegraphics[width=0.47\textwidth]{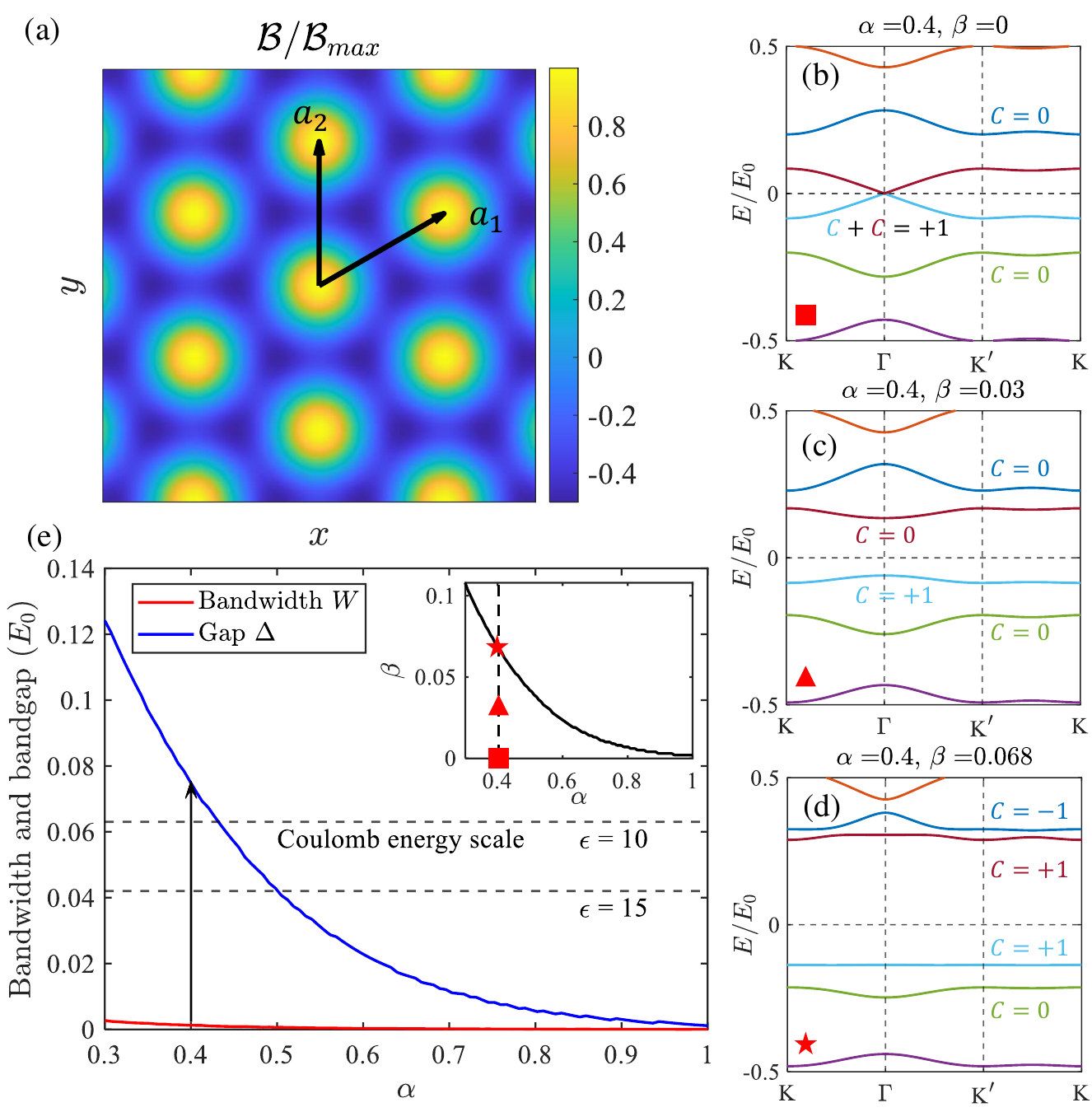}
    \caption{(a) The PMF as described by Eq.~\eqref{PMF}. The two lattice vectors are highlighted. The band structures of two systems having Hamiltonian in Eq.~\eqref{dimensionlessHamiltonian} without (b) and with (c,d) scalar potential. The Chern numbers of the middle four bands are labeled. (e) The minimal bandwidth $W$ of the $C=+1$ flat band below the zero energy and its bandgap $\Delta$ with respect to the lower band for different $\alpha$ by applying the scalar potential that minimizes the bandwidth as shown in the inset. The red square, triangle, and star in the inset label the parameters used in plotting (b-d). The relative energy scale has been emphasized by an arrow for $\alpha=0.4$. All energy scales are measured in units of $E_0 = \hbar v_F |G_0|$. This is equal to $0.3$ eV for the parameters of Ref.~\cite{mao2020evidence}.}
    \label{fig:band_structure}
\end{figure}

\emph{Flat bands and topology}--- Our starting point is the continuum model of strained graphene with a triangular $C_2 \T$-breaking PMF \cite{mao2020evidence} that has the form
\begin{equation}\label{PMF}
    \mathcal{B}(z, \bar z) =\mathcal{B}_0 \sum_{l=0}^5 e^{i \bG_l\cdot \br}= \mathcal{B}_0 \sum_{l=0}^5 e^{\frac{i}{2} (G_l \bar z + \bar G_l z)},
\end{equation}
where $\bG_l=R_{\pi l/3} \bG_0$, $ \bG_0= \frac{4\pi}{\sqrt{3} L_M}(1,0)$ are the 6 smallest reciprocal lattice vectors, and $G_l\equiv G_{lx}+iG_{ly}$.

The Hamiltonian in a single valley has the form $\H = v_F \bsigma \cdot (-i \hbar \boldsymbol{\nabla} + e \widetilde{\boldsymbol{\mathcal{A}}})$ where $\boldsymbol{\nabla} \times \widetilde{\boldsymbol{\mathcal{A}}} = \boldsymbol{\mathcal{B}}$. The other valley is generated by time-reversal symmetry $\T$. $\H$ is invariant under three-fold rotation $C_3$ and $M_x \mathcal{T}$, the combination of  mirror $x \mapsto -x$ and time-reversal.  Strain breaks both $C_2 \T$ and $M_y$ symmetries of graphene \cite{phong2022boundary,de2022network}.
Furthermore, $\H$ has the chiral symmetry $\sigma_z \H \sigma_z = -\H$, which protects a single Dirac cone per valley against gapping out even though $C_2 \T$ symmetry is broken. 
A sublattice potential $\propto \sigma_z$ can be used to open a gap at the Dirac cone, but such a potential cannot be freely tuned in practice.
On the other hand, since the Dirac cone is only protected by chiral symmetry, we can in principle open a gap using a purely scalar potential $\propto \sigma_0$. 
To find such a potential, we note that the sublattice polarized wavefunctions at the Dirac point are given by the simple exponentials $\psi_{A/B} \propto e^{\pm \phi}$ (with $-\nabla^2 \phi \propto \cal{B}$, see Eq.~\ref{Dpsi}) which are peaked at the maxima/minima of $\phi$. Hence, a scalar potential $\propto \phi$ will act as a tunable sublattice potential that opens a gap at the Dirac point. 
The explicit form of the potential is $\H_V = \sigma_0 V_0 \sum_l e^{i \bG_l \cdot \br}$, which matches precisely the PMF pattern. This potential matches the height buckling pattern \cite{mao2020evidence} so it can be generated by a vertical electric field \cite{CorrelationsBuckled,footnote6}. 

It is convenient to express the Hamiltonian in dimensionless units by measuring the momentum in units of $|\bG_0| = \frac{4\pi}{\sqrt{3} L_M}$ and introducing the magnetic length for the PMF $\mathcal{B}_0 = \frac{\hbar}{e l_\mathcal{B}^2}$. Then we can write
\begin{equation}\label{dimensionlessHamiltonian}
    \H = E_0 ([\bk + \alpha \boldsymbol{\mathcal{A}}] \cdot \bsigma + \beta V(
\br)),
\end{equation}
where $ E_0 = \hbar v_F  |\bG_0|$, $ \alpha=1/l_B^2 |\bG_0|^2=3(L_M/4\pi l_B)^2$, and $\beta = V_0/E_0 $ are constants, and $\boldsymbol{\mathcal{A}}$ and $V$ are dimensionless gauge and scalar potentials given by
\begin{equation}\label{Potentials}
\mathcal{A} = \sum_{l=0}^5 e^{i \frac{\pi l}{3}} e^{\frac{i}{2} (G_l \bar z + \bar G_l z)}, \quad V = \sum_{l=0}^5  e^{\frac{i}{2} (G_l \bar z + \bar G_l z)},
\end{equation}
where $\mathcal{A} \equiv \mathcal{A}_x + i \mathcal{A}_y$.
Using the experimental parameters $L_M \approx 15$ nm and $l_B \approx 3.2$ nm, the setup of Ref.~\cite{mao2020evidence} corresponds to $\alpha \approx 0.4$ and $E_0 \approx 0.3$ eV. Fig.~\ref{fig:band_structure}(b-c) show typical band structures  for $\alpha = 0.4$ without ($\beta=0$) and with ($\beta\neq 0$) scalar potentials. For $\beta=0$, the most prominent feature is a pair of isolated bands. They are connected by a single Dirac cone protected by chiral symmetry $\{\H, \sigma_z\} = 0$. $C_3$ symmetry further pins this Dirac cone at the graphene valley ($\Gamma$ point for the supercell). 

To highlight the role of topology, we adopt a sublattice basis. For $\beta = 0$, $\{\sigma_z, \H\} = 0$ which means that $[\sigma_z, \H^2] = 0$, thus we can label the eigenfunctions of $\H^2$ (which are doubly degenerate) by a sublattice index $\psi_{A/B,\bk}$. These wavefunctions are linear superpositions of the energy eigenfunctions $\psi_{A/B,\bk} = (1/\sqrt{2})(\psi_{\epsilon, \bk} \pm \sigma_z \cdot \psi_{\epsilon,\bk})$ where $\sigma_z \cdot \psi_{\epsilon,\bk} \propto \psi_{-\epsilon,\bk}$. Importantly, while the wavefunctions for the lower/upper band around neutrality are singular at the Dirac point and cannot be assigned a Chern number, the sublattice wavefunctions are well-defined everywhere \cite{bultinckGroundStateHidden2020, ledwithStrongCouplingTheory2021, TBGIV}. In the SM \cite{SM}, we show that the sum of these two Chern numbers is necessarily odd, implying that these two bands are non-trivial within a single valley \cite{phong2022boundary}. 
By direct computation, the sublattice $A$ ($B$) wavefunction has Chern number $+1$ (0) in the $K$ valley.

Adding a scalar potential with $\beta >0$ gaps the Dirac point and leads to a well-isolated $C = 1$ band  polarized on the $A$ sublattice as shown in Fig.~\ref{fig:band_structure}(c). Remarkably, the scalar potential can be tuned to obtain an almost perfectly flat band, shown in Fig.~\ref{fig:band_structure}(d). At $\alpha =0.4$, the experimental value in Ref.~\cite{mao2020evidence}, $\beta =0.068$ gives the minimal bandwidth. Using a height modulation around $0.2$ nm \cite{mao2020evidence}, this can be generated by a vertical electric field of 100 mV/nm.

The minimal bandwidth is plotted as a function of $\alpha$ in Fig.~\ref{fig:band_structure}(e) together with the corresponding gap to the closest band with the value of $\beta$ at which this minimum is realized given in the inset. We note that all energy scales decrease exponentially with $\alpha$. 
This exponential squeezing of bands was also observed in chiral TBG for large inverse angle $\alpha$ \cite{tarnopolsky2019origin} and will be explained below. 
On top of the exponential squeezing, the scalar potential further flattens the topological band leading to a minimum bandwidth that is smaller by almost two orders of magnitude relative to the typical energy scale at a given $\alpha$. This almost flat topological band then opens possibilities for exploring strongly correlating physics, which will be discussed below.
For interacting physics, it is instructive at this point to also introduce the scale of the Coulomb interaction: $V_C= e^2/(4\pi \epsilon \epsilon_0 L_M)$.
In dimensionless units, $v_C = V_C/E_0=\sqrt{3} e^2/8\pi^2 \epsilon \epsilon_0 v_F \hbar\approx 0.63/\epsilon$
which is independent of $L_M$. 
In Fig.~\ref{fig:band_structure}(e), we show the energy hierarchy of the bandwidth and the band gap as compared to the Coulomb energy scale. We can see the the bandwidth is significantly smaller than the Coulomb scale, placing the system in the strongly interacting regime.

\begin{figure}
    \centering
    \includegraphics[width=0.44\textwidth]{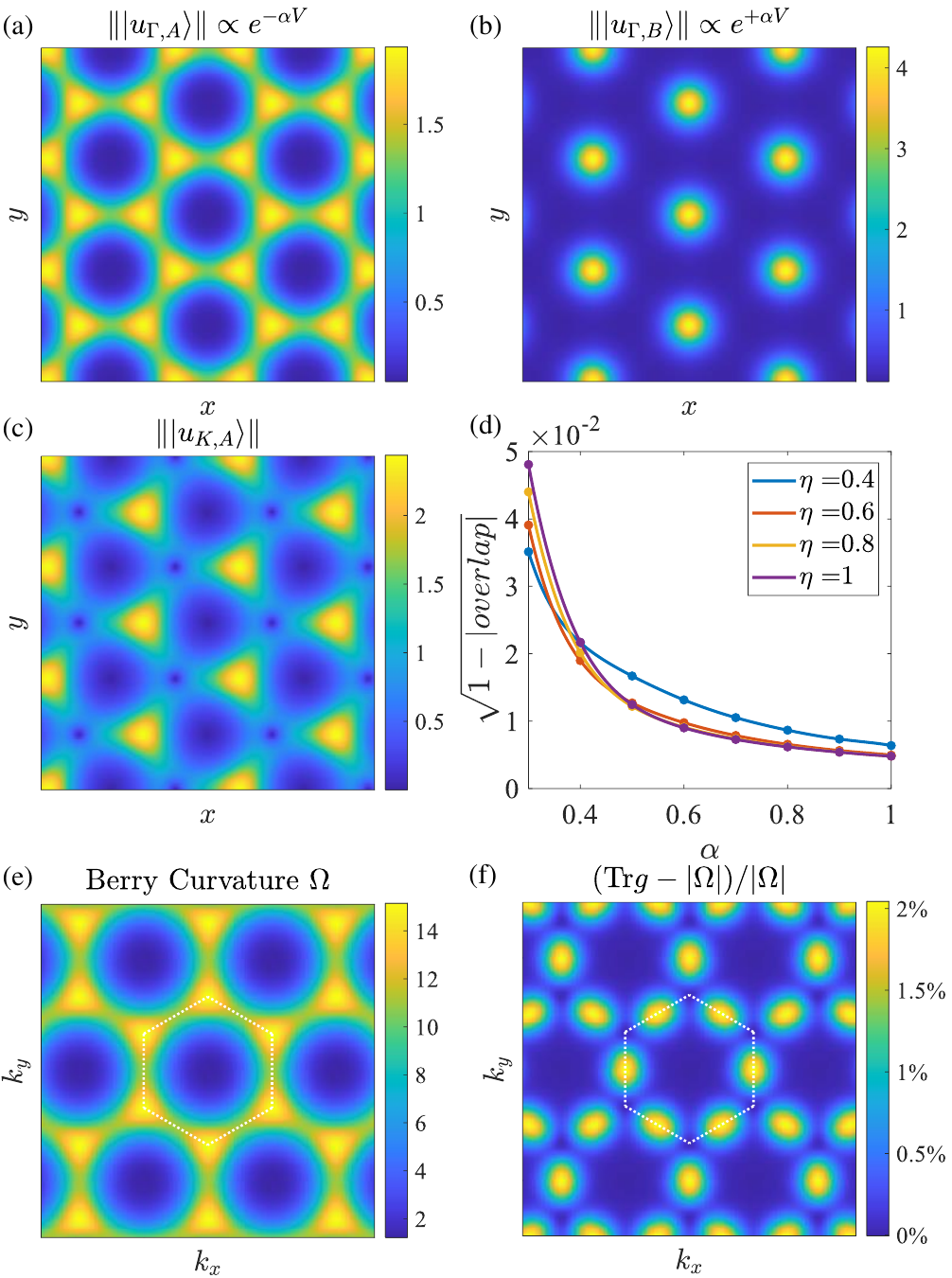}
    \caption{(a,b) Two zero mode wavefunctions at the $\Gamma$ point and (c) the amplitude of the wavefunction at the $K$ point. (d) The square root deviation $\sqrt{1-|\mathrm{overlap}|}$ between the real wavefunction and the ansatz for $\beta = 0$ in Eq.~\eqref{ansatz} averaged over all $\bk$-points in the BZ. (e,f) The Berry curvature $\Omega$ and the trace condition violation $(\text{Tr}g-|\Omega|)/|\Omega|$ of the $C=+1$ band of interest (see main text). Parameters are $\alpha=0.4,\beta=0$. The dotted hexagon indicates the BZ.}
    \label{fig:wave_function}
\end{figure}

\emph{Wavefunctions and quantum geometry}---
We now consider the wavefunctions of the middle two bands for $\beta =0$. We choose to measure the momentum relative to the graphene $K_{g}$ point such that the Bloch wavefunctions are
$\psi_{\bk, A/B}(\br) = e^{i (\bk - K_{g})\cdot \br} u_{\bk, A/B}(\br)$. At the shifted $\Gamma$ point (i.e., $\bk-K_{g}=0$)~\cite{footnote3}, we have a pair of zero modes satisfying the equations
\begin{equation}\label{Dpsi}
    \D \psi_{\Gamma,B} = \D u_{\Gamma,B} = 0, \qquad \D^\dagger \psi_{\Gamma,A} = \D^\dagger u_{\Gamma,A} = 0
\end{equation}
with $\D = -2i \partial + \alpha \bar{\mathcal{A}} $ and $ \D^\dagger = -2i \bar \partial + \alpha \mathcal{A}$.
Noting that $\mathcal{A} = -2i\bar \partial V$, we can easily solve Eq.~\eqref{Dpsi} as $\psi_{\Gamma,A/B}(\br) = e^{\mp \alpha V(\br)}$. These wavefunctions are plotted in Fig.~\ref{fig:wave_function}(a,b). We note that unless $\alpha$ is small, the weight of the A sublattice wavefunction $\psi_{\Gamma,A}$ is strongly suppressed at $\br = 0$ and peaked at the two other $C_3$ invariant points which are related by $M_x \T$, while the B sublattice wavefunction $u_{\Gamma,B}$ is strongly peaked at $\br = 0$. Meanwhile, the wavefunction at the $K$ point for the A sublattice polarized state (i.e., topologically nontrivial) is also shown in Fig.~\ref{fig:wave_function}(c).

To understand the quantum geometry of the bands, let us review the argument of Ref.~\cite{tarnopolsky2019origin} which showed that we can construct an ideal perfectly flat Chern band for a Dirac operator if the zero mode wavefunction at the Dirac point $\psi_0$ has a zero somewhere in real space. Assuming the zero is at $\br = 0$ to be compatible with rotation symmetry, the wavefunctions take the form
\begin{equation}\label{initial_ansatz}
    \psi_\bk(\br) = \frac{\sigma(z + i B^{-1} k)}{\sigma(z)} e^{\frac{i}{2} z \bar k} \psi_0(\br),
\end{equation}
where $k=k_x+ik_y$ and $B = \frac{2\pi}{A_{\rm UC}} = \frac{A_{\rm BZ}}{2\pi}$ with $A_{\rm UC}$ and $A_{\rm BZ}$ being the areas of the unit cell and the Brillouin zone (BZ), respectively.
These wavefunctions satisfy $\D(\bar \partial) \psi_\bk = 0$ if $\D(\bar \partial) \psi_0 = 0$ and transform as Bloch states under translations $\psi_\bk(\br + \bR) = e^{i \bk \cdot \bR} \psi_\bk(\br)$ for any lattice vector $\bR$. The latter property follows from the properties of the modified Weierstrass sigma function \cite{WeierstrassHaldane,wangExactLandauLevel2021b} which has a zero at $z = 0$ and satisfies $\sigma(z + R) = \eta_\bR e^{\frac{B}{2} \bar R (z + R/2)} \sigma(z)$ where $\eta_\bR = +1$ if $\bR/2$ is a lattice vector and $-1$ otherwise. 

The wavefunctions \eqref{initial_ansatz} host ideal quantum geometry in a specific sense that we now describe. A crucial property of the wavefunction (\ref{initial_ansatz}) is that its cell-periodic part $u_\bk = e^{-i \bk \cdot \br} \psi_\bk$ is a holomorphic function of $k$. This property is equivalent \cite{MeraOzawa, OzawaMeta, vortexability} to the trace condition, $\tr g(\bk) = |\Omega(\bk)|$ where $g_{\mu \nu}(\bk)$ is the Fubini-study metric, defined as the symmetric part of the quantum metric tensor $\eta_{\mu \nu}(\bk) = \langle \partial_{k_\mu} u_\bk|(1 - |u_\bk\rangle\langle u_\bk|)| \partial_{k_\nu} u_\bk \rangle$, and $\Omega(\bk)$ is the Berry curvature. Equivalently, this property has been recently interpreted as a vortex attachment condition, which enables the construction of trial FCI states that are guaranteed to be exact ground states for repulsive short-range interactions \cite{vortexability, ledwithFamilyIdealChern2022, HigherChern}. These three equivalent properties define an ideal band. 

Since the wavefunction $\psi_{\Gamma,A}$ is given by a simple exponential, it cannot have any zeros. However, for $\alpha$ sufficiently large \cite{footnote8}, this wavefunction is exponentially small at $\br = 0$. As a result, we can multiply it by a regulator $f_\eta(\br)$ which vanishes at 0 but is close to 1 everywhere else; such a replacement will only change the wavefunction by an exponentially small term. We further require the wavefunction to be rotationally symmetric, which means that it can only depend on $|\br|$. One possible choice of regulator is $f_\eta(\br) = 1-e^{-\eta\br^2}$ for some $\bk$-independent $\eta >0 $. Define an (unnormalized) variational state
\begin{equation}\label{ansatz}
    \psi^\eta_{\bk,A}(\br) = \frac{\sigma(z + i B k)}{\sigma(z)} e^{\frac{i}{2}z \bar k} f_\eta(\br) e^{-\alpha V(\br)},
\end{equation}
whose Bloch periodic part $u^\eta_{\bk,A}= e^{-i\bk\cdot\br}\psi^\eta_{\bk,A}$ is a holomorphic function of $k$, meaning that this ansatz satisfies the ideal band condition. Thus, the deviation of the real wavefunction from the ansatz provides a measure for the violation of the ideal band condition. This deviation, measured by $\sqrt{1 - |\langle \psi_{\bk,A}|\psi_{\bk,A}^\eta\rangle|}$\cite{footnote4} is plotted in Fig.~\ref{fig:wave_function}(d) for different values of $\eta$.
The error decreases with $\alpha$, as expected, and is always $<0.3\%$. This indicates that the violation of the trace condition $\mathrm{Tr}g-|\Omega|$ is very small [see Fig.~\ref{fig:wave_function}(e,f)].
The trace violation is further reduced when $\beta$ is tuned to give the minimal bandwidth (see S.M.~\cite{SM}). We note that the wavefunction \eqref{ansatz}, up to a $\bk$-independent phase, corresponds to the LLL of a Dirac particle in an inhomogeneous magnetic field ${\mathcal B}(\br) = -\nabla^2 \log  \abs{f_\eta(\br) e^{-\alpha V(\br)}/\sigma(z)}$ that has a \emph{non-zero} average flux of $2\pi$ per unit cell \cite{ledwithFractionalChernInsulator2020a}.

The wavefunction of the B sublattice, which is topologically trivial and Wannierizable, is strongly peaked at $\br = 0$. Thus, we can write an ansatz~\cite{footnote5} for the Bloch wavefunction at any $\bk$ given by $\psi_{\bk,B}(\br) = \sum_\bR e^{i \bk \cdot \bR} e^{\alpha V'(\br - \bR)}$ where $V'(\br) = V(\br)$ for $\br$ within the unit cell centered at 0 and $-\infty$ otherwise. Combined with the ansatz for the sublattice A wavefunction, Eq.~\eqref{ansatz}, we see that projecting the $\beta = 0$ Hamiltonian onto the two flat bands yields exponentially small dispersion; the Hamiltonian only contains sublattice off-diagonal terms which contain the overlaps  $\langle\psi_A|\psi_B \rangle \sim e^{-\alpha}$. This also explains why the value of the scalar potential $\beta$ needed to flatten the band decreases exponentially with $\alpha$ [cf.~ the inset in Fig.~\ref{fig:band_structure}(e)]. A detailed analysis of the band energetics is provided in SM~\cite{SM}.

\begin{figure}
    \centering
    \includegraphics[width=0.47\textwidth]{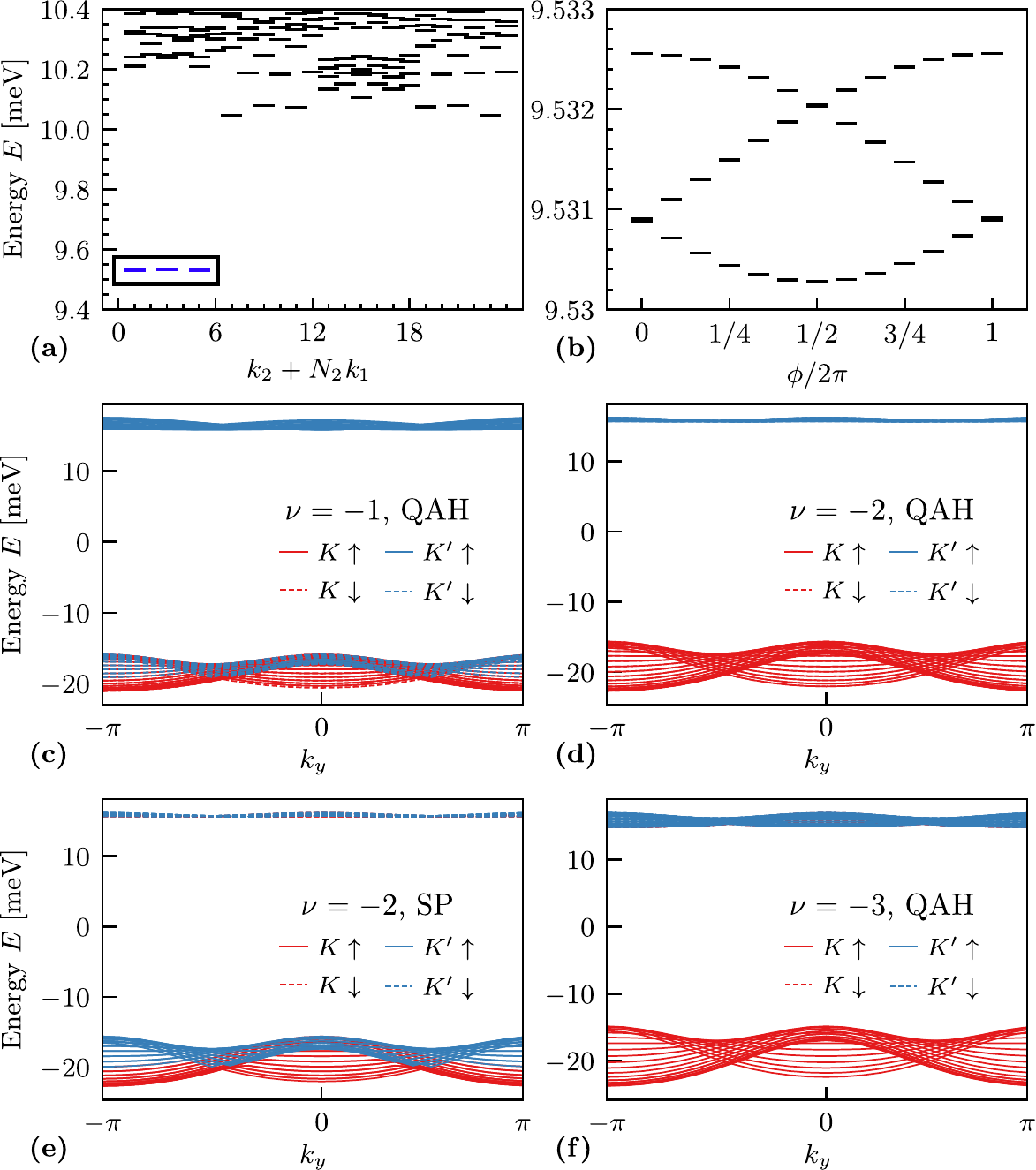}
    \caption{(a) Exact diagonalization spectrum at $\nu=-2/3$ on $24$ $k$-points of the QAH band at $\nu=-1$, as discussed in the text. The ground state is approximately three-fold degenerate (box).
    (b) Spectral flow of the three ground states under flux insertion, indicating a Laughlin state.
    (c-f) Self-consistent Hartree Fock spectra of the strongly-correlated insulators discussed in the text. System size $24 \times 24$. Parameters: $\alpha=0.4,\beta=0.068$, and $E_0 = 0.325$ eV \cite{footnote7}
    }
    \label{fig:ED_HF}
\end{figure}

\emph{Interacting phases for the partially filled Chern band}--- Next we consider the effect of interactions upon partially filling the flat Chern band by hole doping the band structure in Fig.~\ref{fig:band_structure} relative to charge neutrality. Due to valley and spin, we consider the filling $\nu \in [-4,0]$. Using a screened Coulomb interaction $V_\bq = \frac{e^2}{2 \epsilon \epsilon_0 |\bq|} \tanh |\bq| d$, we consider an interacting Hamiltonian $\mathcal{H} + \mathcal{H}_{\rm int}$ with \cite{bultinckGroundStateHidden2020, ledwithStrongCouplingTheory2021}
\begin{equation}
    \H_{\rm int} = \frac{1}{2A} \sum_\bq V_\bq \delta \rho_\bq \delta \rho_{-\bq}, \quad \rho_\bq = \sum_{\alpha,\bk} \lambda_{\alpha,\bq}(\bk) c^\dagger_{\alpha,\bk} c_{\alpha,\bk + \bq} 
\end{equation}
where $\delta \rho_\bq = \rho_\bq - \bar \rho_\bq$, $\bar \rho_\bq = \sum_{\alpha,\bG,\bk} \delta_{\bq,\bG} \lambda_{\alpha,\bG}(\bk)$. Here, $\alpha = (s, \tau)$ is a combined index for spin $s$ and valley $\tau$, $\bG$ are reciprocal lattice vectors, and $\lambda_{\alpha,\bq}(\bk) = \langle u_{\alpha,\bk}| u_{\alpha,\bk + \bq}\rangle$ are form factors. 

In the limit of small bandwidth, we can employ strong coupling analysis similar to that done in TBG \cite{KangVafekPRL, bultinckGroundStateHidden2020, ledwithStrongCouplingTheory2021, TBGIV} to deduce that the ground states at integer fillings are generalized spin-valley ferromagnets. The argument is explained in detail in SM \cite{SM} and summarized here. Our setup is simpler than TBG, since there is a single band per spin and valley. It is also simpler than other moir\'e systems like twisted double bilayer graphene which have a single band per spin and valley but whose dispersion is non-neglegible \cite{leeTheoryCorrelatedInsulating2019}. At $\nu = -1$ and $\nu = -3$, the ground state is a QAH spin and valley polarized insulator with Chern number $\pm 1$ that spontaneously breaks both $\SU(2)$ spin and time-reversal $\T$. At $\nu = -2$, we have two degenerate ground state manifolds: (i) a QAH valley ferromagnet with $C = \pm 2$ and (ii) a family of spin-polarized states with $C = 0$ consisting of a spin ferromagnet in each valley. The two manifolds (i) and (ii) are degenerate in our model, but adding an intervalley Hund's coupling is expected to lift the degeneracy and select states in (ii) \cite{SM,YahuiFlatChern, leeTheoryCorrelatedInsulating2019}.

In contrast to TBG, there are no further anisotropies to consider here. In addition, intervalley coherent orders are strongly disfavored since they involve coherent superposition of states from opposite Chern bands, leading to nodes in the order parameters \cite{bultinckMechanismAnomalousHall2020, leeTheoryCorrelatedInsulating2019}. Furthermore, the interaction-generated dispersion due to Hartree-Fock corrections \cite{KangVafekPRL, RepellinYahui, TBGV, KangBernevigVafek} is smaller compared to TBG with similar interaction parameters \cite{SM}. This follows from the delocalization of the $A$-sublattice wavefunctions across two different points, related by $M_x \T$, (see Fig. \ref{fig:wave_function}a) which leads to a much milder Hartree potential than that of the AA-site-localized TBG electrons. This makes the QAH more energetically favored against competing states compared to TBG \cite{pierce2021unconventional}. The ground states at different fillings are confirmed through self-consistent Hartree-Fock,  shown in Fig.~\ref{fig:ED_HF}, which verify the QAH states at $\nu = -1$, and $-3$ and the degenerate spin and valley polarized states at $\nu = -2$. We notice here the relatively large gaps and small quasiparticle dispersion compared to TBG (see SM \cite{SM} for comparison).

We expect that the flat ideal Chern bands of the A sublattice will host FCIs when fractionally filled. We verify this in the simplest case where we electron-dope the $\nu = -1$ spin and valley polarized QAH state, such that the doped charge enters in a single flavor. We study the filling $\nu = -2/3$ using single-flavor exact diagonalization and show our results in Fig.~\ref{fig:ED_HF}.
 We see clear signatures of a Laughlin state with 3-fold ground state degeneracy and spectral flow indicating topological order. We note that we have not included the interaction-generated dispersion. Includng this introduces inhomogenieties that make ED extremely sensitive to grid choice. We note however the DMRG results of Ref.~\cite{parkerFieldtunedZerofieldFractional2021} showed that FCIs in chiral TBG are stable up to relatively large values of dispersion. Given the milder Hartree dip in our setup that makes the interaction-generated dispersion a lot smaller compared to TBG \cite{SM}, we expect the FCIs to survive its addition.
We leave a detailed analysis of this effect to future works.

\emph{Discussion}--- We studied a system of monolayer graphene with periodic, $C_2$-breaking pseudo-magnetic field combined with a periodic scalar field with the same period $L_M \gg a$. This system may be realized experimentally by placing graphene on top of a $C_2$-breaking substrate such as NbSe${}_2$. 
The substrate causes both strain, leading to a $C_2$-breaking PMF, and height modulation, giving a periodic potential in perpendicular electric field. Other realizations involve a network of nanorods \cite{Nanorods} arranged in a $C_2$-breaking pattern (see Ref.~\cite{phong2022boundary}), combined with a periodic scalar potential (which can be generated by a patterned dielectric \cite{CanoPixleyMoireSurface, CanoPixleyMeronLattice} or a separate moir\'e hBN potential \cite{MoirehBN}).  
Our analysis has shown that this system is simpler and more tunable than most graphene-based moir\'e systems even in ideal limits. Thus, it represents an extremely promising platform to realize quantum anomalous Hall states and fractional Chern insulators, as we have shown numerically. Furthermore, by switching the sign of the scalar field or the gate voltage, we can access both a topological band and a trivial band within the same system. From an experimental viewpoint, the main technical challenge in the current setup based on NbSe${}_2$ substrate lies in the difficulty of gating the sample since the substrate is metallic. By overcoming this technical difficulty or using a different $C_2$-breaking but insulating substrate, we predict this system to be an ideal platform to study strong correlation effects in topological bands with significant advantages over twisted multilayer graphene-based moir\'e systems.

\begin{acknowledgements}
{\it Acknowledgements}--- We thank Ashvin Vishwanath for helpful discussions and collaborations on related topics. Q.G. acknowledges the support of the Provost’s Graduate Excellence Fellowship from the University of Texas at Austin. P.J.L. was supported by the Department of Defense (DoD) through the National Defense Science and Engineering Graduate Fellowship (NDSEG) Program. This research is funded in part by the Gordon and Betty Moore Foundation’s EPiQSInitiative, Grant GBMF8683 to D.E.P.
\end{acknowledgements}

\bibliographystyle{unsrt}
\bibliography{Reference}

\begin{thebibliography}{100}
\bibitem{footnote6}
{We note that although a $\sigma_z$ term is symmetry-allowed, its effect on the
  two bands close to neutrality can be absorbed into the scalar potential which
  provides an approximately constant $\sigma_z$ term \cite{SM}.}
\bibitem{footnote3}
{We use capital letters with subscript `g' to denote the graphene high
  symmetric $k$-points and capital letters without subscript the high symmetric
  $k$-points of the supercell, the latter of which will be shifted by the
  graphene valleys $K_g$.}
\bibitem{footnote8}
{While the value of $\alpha$ we use, $\alpha = 0.4$ is not large, we note that
  in our convention, the maximum value of the potential is 6 which leads to the
  exponential factor $e^{-0.4*6} = e^{-2.4} \approx 0.09$}.
\bibitem{footnote4}
{Let us write the real wavefunction for $A$ sublattice at $\boldsymbol{k}$ as
  the ansatz plus a small deviation: $|\psi_{\boldsymbol{k},A}\rangle =
  |\psi_{\boldsymbol{k},A}^{\eta}\rangle+|\delta\psi_{\boldsymbol{k},A}\rangle$.
  If the real wavefunction and ansatz are both normalized, we have $ \vert|
  |\psi_{\boldsymbol{k},A}\rangle\vert| ^2 = 1 =1 + 2\text{Re} \langle \delta
  \psi_{\boldsymbol{k},A}| \psi_{\boldsymbol{k},A}^{\eta}\rangle + \vert|
  |\delta \psi_{\boldsymbol{k},A}\rangle\vert| ^2$. Thus, to the first order in
  $|\delta\psi_{\boldsymbol{k},A}\rangle $, we have $ \text{Re} \langle \delta
  \psi_{\boldsymbol{k},A}| \psi_{\boldsymbol{k},A}^{\eta}\rangle=0$, which
  indicates that overlap deviation $1-| \langle\psi_{\boldsymbol{k},A}|
  \psi_{\boldsymbol{k},A}^{\eta}\rangle|$ goes quadratic with
  $|\delta\psi_{\boldsymbol{k},A}\rangle $. However, the trace condition
  violation normally goes linearly with $|\delta\psi_{\boldsymbol{k},A}\rangle
  $. Thus, we plot the square root overlap deviation in
  Fig.~\ref{fig:wave_function}(d) as to better compare with trace violation.}
\bibitem{footnote5}
{The reader may wonder why a similar ansatz to that of sublattice $A$ does not
  work here, which would yield a $C=-1$ band antiholomorphic in $k_x + i k_y$.
  In fact, such an ansatz is possible but it necessarily mixes multiple
  (exponentially-squeezed) $B$-sublattice bands. However, we expect the
  topologically-trivial strongly-localized band, that we obtain from
  diagonalzing the Hamiltonian, to dominate the low-energy $B$-sublattice
  physics because it is straightforward to keep particles apart in this band.
  In contrast, the $A$ sublattice bands have a more robust topology because
  their density is localized at two distinct points in the unit cell related by
  $M_x\T$.}
\bibitem{footnote7}
{This value corresponds to $L_M \approx 13.3$ nm which matches that of TBG at
  the first magic angle. This makes it easier to compare the resulting gaps and
  dispersion with those of TBG.}
\end{thebibliography}

\renewcommand{\theequation}{S\arabic{equation}}
\setcounter{equation}{0}
\onecolumngrid
\section{Supplemental Material}
This Supplementary Material contains detailed discussions on symmetries of the Hamiltonian, the effect of the scalar potential, band energetics, and interacting phases.
\vspace{12 pt}
\twocolumngrid
\subsection{Symmetries of the Hamiltonian}
In this section, we briefly overview the symmetries of the Hamiltonian (\ref{dimensionlessHamiltonian}) with the gauge field $A$ and scalar field $V$ given by Eq.~\ref{Potentials}. In particular, the $x$ and $y$ components of the field are given by
\begin{gather}
     A_x = \Re A = i \sum_l \sin \frac{\pi l}{3} e^{i \bG_l \cdot \br} = i \sum G_{l,y} e^{i \bG_l \cdot \br}, \\ A_y = \Im A = -i\sum_l \cos \frac{\pi l}{3} e^{i \bG_l \cdot \br} = -i\sum_l G_{l,x} e^{i \bG_l \cdot \br},
\end{gather}
where we used $G_l = (\cos \frac{\pi l}{3}, \sin \frac{\pi l}{3})$.

We now focus on the action of symmetries on the single valley Hamiltonian. The other valley is generated by the action of time-reversal symmetry. In the absense of any gauge fields (i.e. strain), the single-valley Dirac Hamiltonian $\H_D(\br) = -i \bsigma \cdot \nabla$ has the following symmetries
\begin{gather}
    C_3 \H_D(\br) C_3^{-1} = e^{-\frac{i \pi }{3} \sigma_z} \H_D(e^{\frac{2\pi i}{3} \sigma_y} \br) e^{i\frac{\pi }{3} \sigma_z}, \\ M_y \H_D(x,y) M_y^{-1} = \sigma_x \H_D(x,-y) \sigma_x \\
    (C_2 \T) \H_D(\br) (C_2 \T)^{-1} = \sigma_x \H^*_D(-\br) \sigma_x, \\ (M_x \T) \H_D(x,y) (M_x \T)^{-1} = \H^*_D(-x,y)
\end{gather}
The vector potential piece is 6-fold rotationally symmetric since $A(z e^{\frac{2\pi i}{6}}) = e^{\frac{2\pi i}{6}} A(z)$, i.e. $A(z)$ transforms as a vector under rotation. Since $A_x(-x,y) = A_x(x,y)$, $A_x(x,-y) = -A_x(x,y)$ and $A_y(-x,y) = -A_y(x,y)$, $A_y(x,-y) = A_y(x,y)$, we find
\begin{gather}
    (C_2 \T) \bA(\br) \cdot \bsigma (C_2 \T)^{-1} = \bA(-\br) \cdot \bsigma = -\bA(\br) \cdot \bsigma \\
    M_y \bA(x,y) \cdot \bsigma M_y^{-1} = A_x(x,-y) \sigma_x - A_y(x,-y) \sigma_y \nonumber \\ = -\bA(\br) \cdot \bsigma \\
    (M_x \T) \bA(\br) \cdot \bsigma (M_x \T)^{-1} = A_x(-x,y) \sigma_x - A_y(-x,y) \sigma_y \nonumber \\ = \bA(\br) \cdot \bsigma
\end{gather}
Thus, the strain field breaks both $C_2 \T$ and $M_y$ while preserving $M_x \T$ and $C_3$. 

\subsection{Band topology}
Here we will show that the total Chern number of the two bands around neutrality is odd. To understand the properties of these bands in the sublattice basis, we note that they are adiabatically connected to the bands obtained by adding to the Hamiltonian a large mass term $\Delta \sigma_z$ projected to the space of the two flat bands. For positive/negative $\Delta$, the lower band is polarized on sublattice $B/A$. For $\sgn \Delta = \gamma = \pm 1$, let us define the Chern number of the $A/B$ sublattice bands to be $C_{A/B,\gamma}.$ The Hamiltonian $\H(\Delta) = \bsigma \cdot (\bk + \bA) + \Delta \sigma_z$ satisfies $\H(\Delta) = -\sigma_z \H(-\Delta) \sigma_z$ which means that $C_{A/B,+} = C_{A/B,-} = C_{A/B}$. On the other hand, for small $\Delta$, we know that the Chern number changes by $+1$ as we change the sign of $\Delta$ from negative to positive (since the Chern number of a continuum Dirac cone is $\frac{1}{2} \sgn (\Delta)$. Thus, $C_{B,+} = C_{A,-} + 1$ which implies $C_{B} = C_A + 1$. As a result, the total Chern number $C = C_A + C_B = 2C_A + 1$ is necessarily odd. This is verified in Fig. \ref{fig:band_structure} (b,c,d) by direct computation.

\subsection{The effect of scalar potential on trace condition violation and bandwidth of the flat $C=+1$ band}
\begin{figure}
    \centering
    \includegraphics[width=0.47\textwidth]{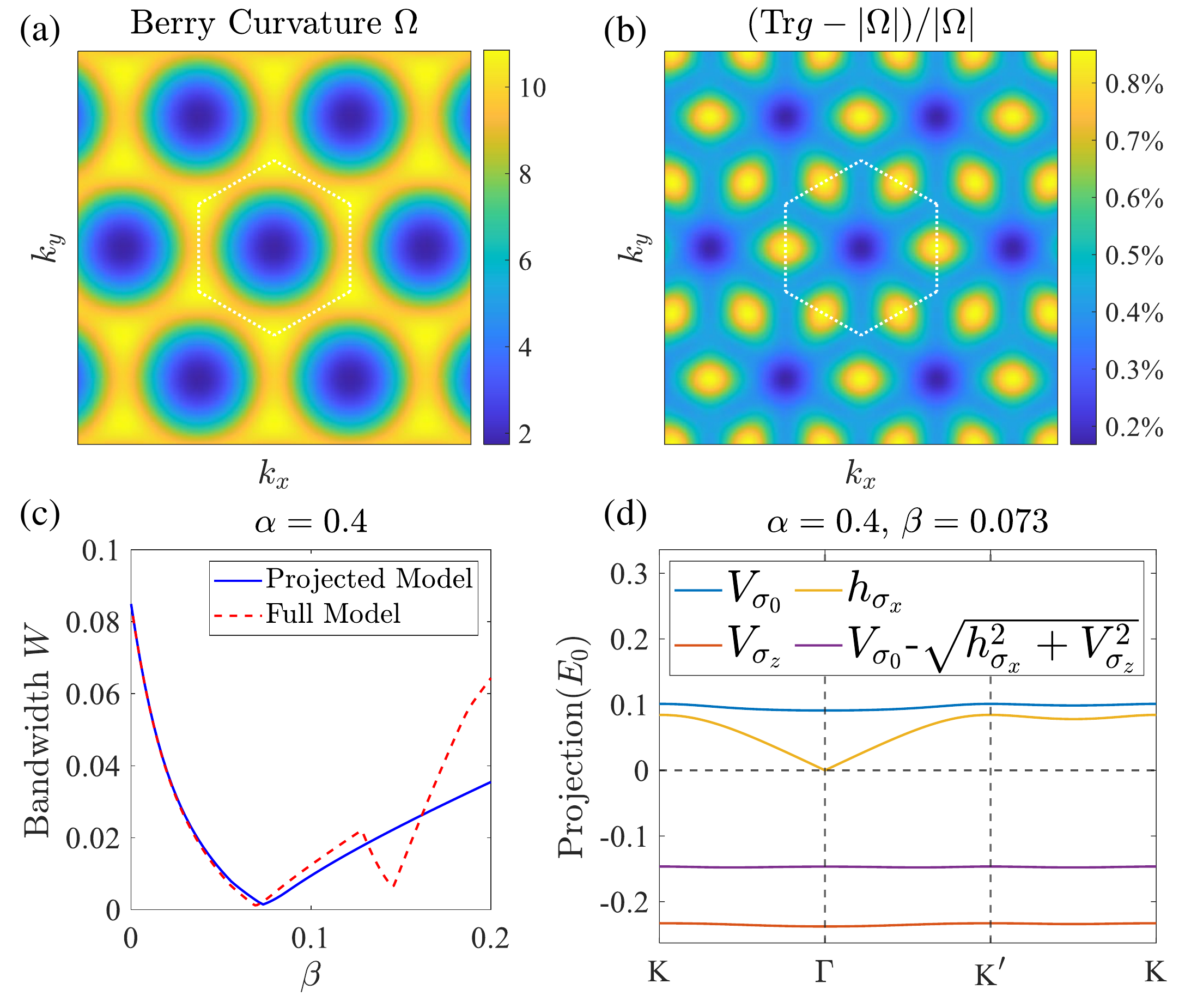}
    \caption{(a) The Berry curvature and (b) the trace condition violation for the system with $\alpha=0.4$ and $\beta = 0.068$. (c) The bandwidth of the flat $C=+1$ band achieved for $\alpha=0.4$ at different $\beta$'s using the full model and projected band model. (d) The parameters from Eqs.\eqref{eq:H0_projected}, \eqref{eq:V_projected} obtained by projecting the Hamiltonian onto the middle two bands in Fig.~\ref{fig:band_structure}(b). The purple curve shows the resulting flat band in the projected model.}
    \label{fig:effect_of_scalar_potential}
\end{figure}

As discussed in the main text, the trace condition for the topological $C=+1$ band is only slightly violated because its wavefunctions are exponentially close to a set of ansatz that satisfies exactly the trace condition. Here we numerically confirm that after the introduction of the scalar potential, the trace condition violation becomes even smaller when the minimal bandwidth of that $C=+1$ band is realized, i.e., $\beta=0.068$ for $\alpha=0.4$ [see Fig.~\ref{fig:effect_of_scalar_potential}(a,b)]. A naive intuition for this extra reduction of the violation is that the scalar potential puts an energy barrier for $A$ sublattice wavefunctions around $\br=0$, making $\psi_{A,\bk=0}(\br=0)$ even closer to zero (recall that an exact zero of $\psi_{A,\bk=0}$ at $\br=0$ implies a flat and ideal band for the $\beta=0$ chiral model). However, for sufficiently large $\beta$ the breaking of chiral symmetry becomes significant enough to increase both the bandwidth and the trace condition violation substantially. By diagonalizing the full model (red dashed curve in Fig.~\ref{fig:effect_of_scalar_potential}(c)), we see that there is an optimal value of $\beta$ for a given $\alpha$ (we show the bandwidth, but the trace condition is also optimized at a nearby value of $\beta$).

To see why there is an optimal value for $\beta$ for optimizing the bandwidth, we project the Hamiltonian onto the middle two bands depicted in Fig.~\ref{fig:band_structure}(c) (i.e., $\alpha=0.4,\beta=0$) using the sublattice polarized wavefunctions $\{ |\psi_{A,\bk}\rangle, |\psi_{B,\bk}\rangle\}$. This projected model is accurate for small $\beta$ and captures the existence of an optimal value as shown in the blue line of Fig.~\ref{fig:effect_of_scalar_potential}(c).
We obtain
\begin{equation}
    \H_0^{2\times 2}(\bk) = \langle \psi_{s,\bk}| [\bk + \alpha \tilde \bA(\br)] \cdot \bsigma|\psi_{s',\bk}\rangle = h_{\sigma_x}(\bk)\sigma_x.
    \label{eq:H0_projected}
\end{equation}
Note that there is also generically a term proportional to $\sigma_y$, but for convenience we choose to rotate it into $\sigma_x$ for each $\bk$ with a (in general singular) sublattice-band-dependent gauge transformation $\psi_{\bk} \to e^{i \theta_\bk \sigma_z} \psi_{\bk}$. Only the gauge invariant squared dispersion will enter below, however.

We also have
\begin{equation}
    V^{2\times 2}(\bk) = \langle \psi_{s,\bk}| \beta V(\br)|\psi_{s',\bk}\rangle = V_{\sigma_0}(\bk)\sigma_0 + V_{\sigma_z}(\bk)\sigma_z,
    \label{eq:V_projected}
\end{equation}
where $s,s'=A,B$. The projected parameters $h_{\sigma_x}$, $V_{\sigma_0} $, and $ V_{\sigma_z}$ are plotted in Fig.~\ref{fig:effect_of_scalar_potential}(d), where one should notice that the shape of $h_{\sigma_x}$ reproduces the dispersion of the middle two bands in Fig.~\ref{fig:band_structure}(b), as expected. 

The final dispersion of this projected model is then $V_{\sigma_0}\pm \sqrt{h_{\sigma_x}^2+V_{\sigma_z}^2} $. 
We note that $V_{\sigma_{0,z}}(\bk) \propto \beta$ has weak $\bk$ dependence, while $h_{\sigma_x}(\bk)$ has a prominent dip at the $\Gamma$ point corresponding to the zero energy Dirac point of the $\beta = 0$ model (see Fig. \ref{fig:effect_of_scalar_potential}(d)). The term $V_{\sigma_z}$ then acts as a mass for the Dirac cone dispersion of $h_{\sigma_{x}}(\bk)$, which yields a rapidly flattening quadratic dispersion $\propto 1/\beta$ as seen in Fig. \ref{fig:effect_of_scalar_potential}(c) for small $\beta$. For large $\beta$ the momentum dependence of $V_{\sigma_{0,z}}(\bk)$ dominates the residual quadratic dispersion, such that the dispersion increases with $\beta$, but because the momentum dependence of $V_{\sigma_{0,z}}(\bk)$ is so weak the total dispersion reaches a tiny value before finally increasing.


\subsection{Strong coupling analysis}
We now discuss interactions in the Chern band directly below charge neutrality and will derive the exact many-body ground states in the strong coupling limit. Due to valley and spin, the band has four "flavors". We consider a Hamiltonian with strong repulsive interactions plus the small dispersion in the four flat bands \cite{bultinckGroundStateHidden2020, ledwithStrongCouplingTheory2021}:
\begin{gather}
\label{IntHam}
    \H = \sum_{\alpha,\bk} c_{\alpha,\bk}^\dagger h_{\alpha}(\bk) c_{\alpha,\bk} + \frac{1}{2A} \sum_\bq V_\bq \delta \rho_\bq \delta \rho_{-\bq},\\ \delta \rho_\bq = \rho_\bq - \bar \rho_\bq,
    \quad \bar \rho_\bq = \sum_{\bG \in \Lambda^*} \delta_{\bq, \bG} \sum_{\alpha,\bk} \lambda_{\alpha,\bG}(\bk),\\
    \rho_\bq = \sum_{\alpha,\bk} \lambda_{\alpha,\bq}(\bk) c_{\alpha,\bk}^\dagger c_{\alpha,\bk + \bq},
\end{gather}
where $V_\bq$ is some repulsive interaction, $c_{\alpha,\bk}$ are the annihilation operators for the flat band below charge neutrality labelled by a flavor index $\alpha =(s,\tau) = (\uparrow/\downarrow, K/K')$, $\Lambda^*$ is the reciprocal lattice, and $h_0$ is the non-interacting dispersion. The form factors are defined as $\lambda_{\alpha,\bq}(\bk) \equiv \langle u_{\alpha,\bk}| u_{\alpha,\bk + \bq} \rangle$ and we assume a periodic gauge such that $\lambda_{\alpha,\bq}(\bk + \bG) = \lambda_{\alpha,\bq}(\bk)$ and $c_{\alpha, \bk + \bG} = c_{\alpha,\bk}$ for any reciprocal lattice vector $\bG$. Intuitively, $\delta \rho$ measures the change in density relative to charge neutrality.

This choice of the Hamiltonian (\ref{IntHam}) guarantees that the interaction term vanishes when acting on a state where the four flat bands are fully filled $|\Psi_{\rm full} \rangle = \prod_{\alpha,\bk} c_{\alpha,\bk}^\dagger |0 \rangle$ (where $|0 \rangle$ is the state where the four flat bands are empty). To see this, we note that
\begin{equation}
    c^\dagger_{\alpha,\bk} c_{\alpha,\bk + \bq} |\Psi_{\rm full} \rangle = \begin{cases}
        0 &\text{if } \bq \notin \Lambda^*\\
        |\Psi_{\rm full} \rangle &\text{if } \bq \in \Lambda^*
    \end{cases}
\end{equation}
which implies $\rho_\bq |\psi_{\rm full} \rangle = \bar \rho_\bq |\psi_{\rm full} \rangle$.

We note that we can alternatively write the interaction in terms of the normal-ordered operators such that the interaction term annihilates the state $|0 \rangle$ where the four flat bands are empty:
\begin{gather}
    \H = \sum_{\alpha,\bk} c_{\alpha,\bk}^\dagger \tilde h_{\alpha}(\bk) c_{\alpha,\bk} + \frac{1}{2A} \sum_\bq V_\bq :\rho_\bq \rho_{-\bq}:
\end{gather}
where the single-particle dispersion is now modified as
\begin{equation}
    \tilde h_{\alpha}(\bk) = h_{\alpha}(\bk) + h^F_\alpha(\bk) - 4h^H_\alpha(\bk)
\end{equation}
and $h^H$ and $h^F$ are the Hartree and Fock dispersions defined as
\begin{align}
    h^F_\alpha(\bk) &= \frac{1}{2A} \sum_\bq V_\bq |\lambda_{\alpha,\bq}(\bk)|^2 \\
    h^H_\alpha(\bk) &= \frac{1}{4A} \sum_{\bG \in \Lambda^*} V_\bG \lambda_{\alpha, \bG}(\bk) \sum_{\beta,\bk'} \lambda_{\beta,-\bG}(\bk').
\end{align}
Using time-reversal symmetry and the periodicity of the gauge, we have the identity $\lambda_{s,-\tau,\bG}(\bk) = \lambda^*_{s,-\tau,-\bG}(\bk + \bG) = \lambda^*_{s,-\tau,-\bG}(\bk) = \lambda_{s,\tau,\bG}(-\bk)$, which implies that $\sum_\bk \lambda_{\beta,\bG(\bk)}$ is independent of the flavor $\beta$. Thus, we can rewrite the Hartree potential as
\begin{equation}
    h^H_\alpha(\bk) = \frac{1}{A} \sum_{\bG \in \Lambda^*} V_\bG \lambda_{\alpha, \bG}(\bk) M_\bG, \; M_\bG = \sum_{\bk'} \lambda_{\uparrow,K,-\bG}(\bk').
\end{equation}
The Hamiltonian (\ref{IntHam}) has a $\U(2) \times \U(2)$ symmetry, consisting of $\U(1)$ charge conservation and $\SU(2)$ spin conservation in each valley. 

\begin{figure}
    \centering
    \includegraphics{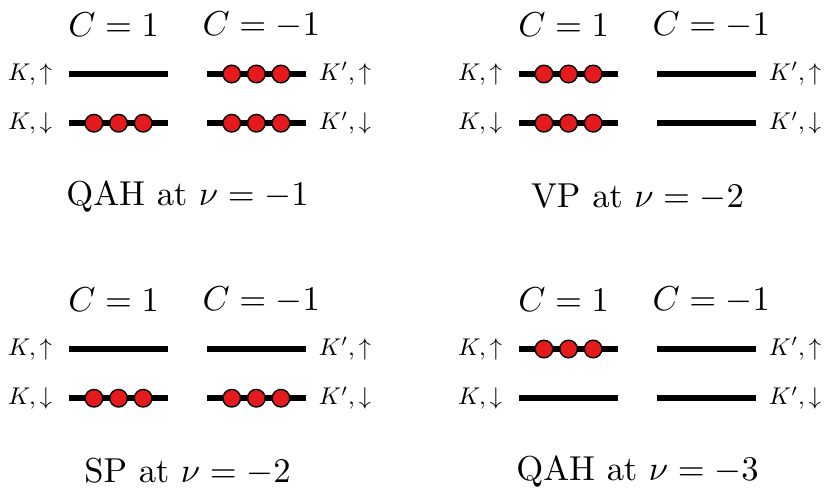}
    \caption{The fillings of the four flavors of flat bands in the various generalized ferrogmangets discussed in the text. Each Chern sector has a $\U(2)\times \U(2)$ symmetry.}
    \label{fig:strong_coupling_diagrams}
\end{figure}

\begin{figure*}
    \centering
    \includegraphics[width = 0.85 \textwidth]{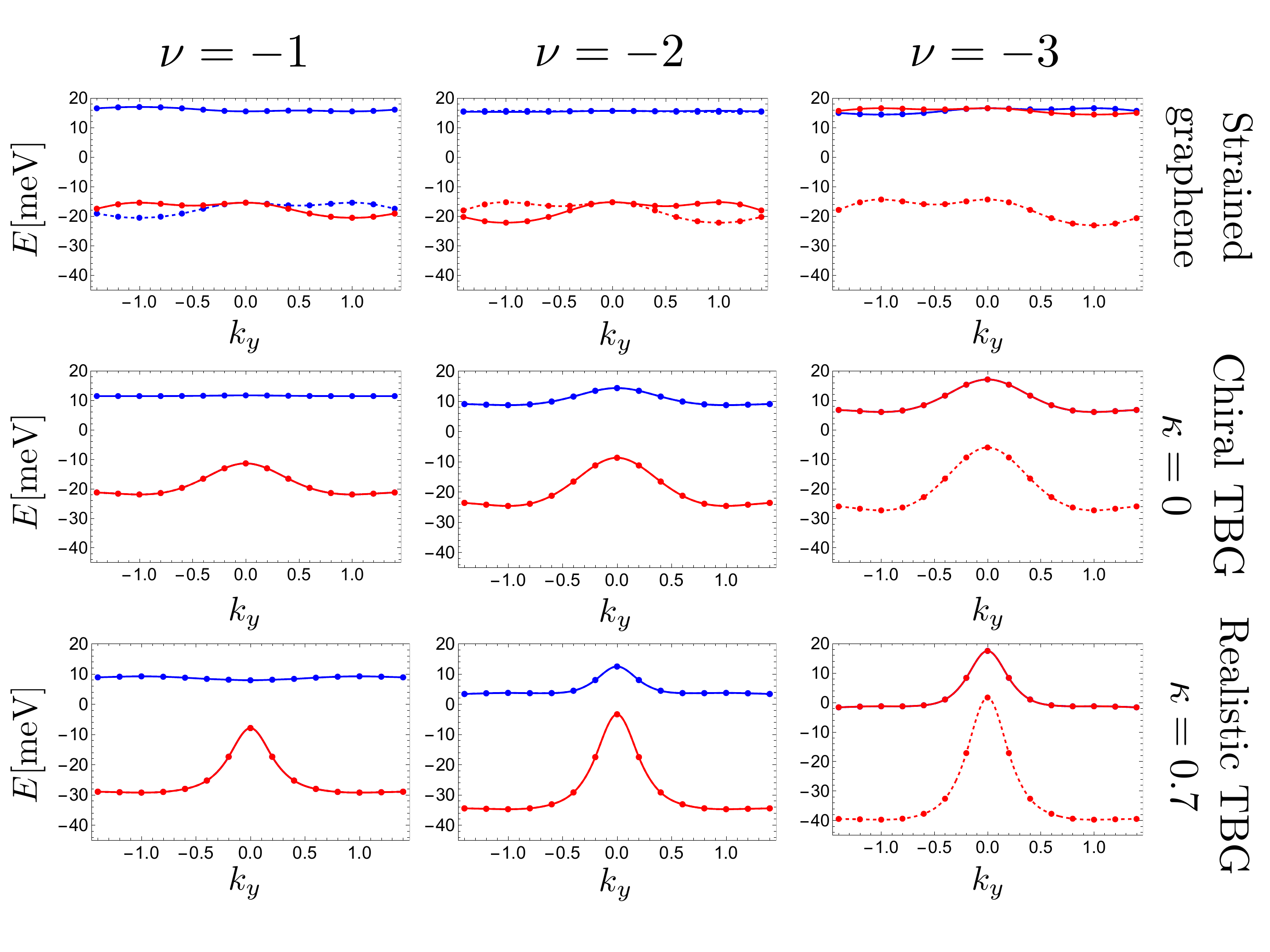}
    \caption{Comparison between our model of periodically strained graphene and hBN-aligned TBG both in the chiral limit $\kappa = 0$ and away from it $\kappa = 0.7$. As in the main text, we used dielectric constant $\epsilon = 10$, screening distance $d = 10$ nm and sublattice potential for the top later $\delta_{\rm top} = 15$ nm \cite{bultinckMechanismAnomalousHall2020}. We used an overall energy scale $E_0 = 0.324$ eV corresponding to the energy scale of TBG at the first magic angle. Detailed values for the gaps and bandwidth are provided in Table \ref{tab:EnergyComparison}.}
    \label{fig:SpectrumComparison}
\end{figure*}

At integer fillings, spin-valley ferromagnets are {\it exact} eigenstates of this Hamiltonian \cite{RepellinYahui, bultinckGroundStateHidden2020, TBGV} in the flat band limit $h_0 = 0$. Fig. \ref{fig:strong_coupling_diagrams} illustrates a few of these states.  The arguments we use below are akin to those used in TBG (see \cite{ledwithStrongCouplingTheory2021} for a pedagogical treatment), but generally simpler as there are half as many flat bands here. To see that flavor ferromagnets are exact eigenstates of the Hamiltonian, we note that the density operator consists of a sum over flavors $\rho_\bq = \sum_\alpha \rho_{\alpha,\bq}$, $\rho_{\alpha,\bq} = \sum_{\alpha,\bk} \lambda_{\bq}(\bk) c_{\alpha,\bk}^\dagger c_{\alpha,\bk + \bq}$ where the density operator only acts within this flavor. Denoting the state where the $\alpha$ flavor is fully filled (empty) by $|\Psi^\alpha_{\rm full} \rangle$ ($|\Psi^\alpha_{\rm empty} \rangle$), we find $\rho_{\alpha,\bq} |\Psi^\alpha_{\rm full} \rangle = \bar \rho_{\alpha,\bq} |\Psi^\alpha_{\rm full} \rangle$ and $\rho_{\alpha,\bq} |\Psi^\alpha_{\rm empty} \rangle = 0$. Thus, for a generalized Ferromagnet (GFM) where each flavor is either full or empty, we find
\begin{gather}
    \H_{\rm int} |\Psi_{\rm GFM} \rangle = E_{\rm GFM} |\Psi_{\rm GFM} \rangle \\
    E_{\rm GFM} = \sum_{\bk, \alpha \,\, {\rm empty}} h_\alpha^H(\bk) = \frac{|\nu|}{A} \sum_{\bG \in \Lambda^*} V_\bG M_{-\bG} M_\bG
\end{gather}
The single particle excitations on top of $|\Psi_{\rm GFM} \rangle$ can be obtained exactly as follows. Using Koopman's Theorem, electron (hole) excitations are obtained as $c_{\alpha,\bk}^\dagger |\Psi_{\rm GFM} \rangle$ ($c_{\alpha,\bk} |\Psi_{\rm GFM} \rangle$) where flavor $\alpha$ is empty (full). The corresponding spectrum is given by
\begin{align}
    \epsilon_{\alpha,e}(\bk) &= + h_\alpha^F(\bk) - |\nu| h_\alpha^H(\bk), \\
    \epsilon_{\alpha,h}(\bk) &= -h_\alpha^F(\bk) - |\nu| h_\alpha^H(\bk).
\end{align}
The GFMs will be the actual ground state of the system if the gap to charged excitation is much larger the quasiparticle dispersion \cite{TBGIV, TBGV} which is the case for our system (see Fig.~\ref{fig:SpectrumComparison}). 

The manifold of GFM ground states at different fillings can be generated by acting with $\U(2) \times \U(2)$ symmetry on simple GFM states. Such a manifold can be concisely parameterized in terms of the correlation matrix $P_{\alpha \beta}(\bk) = \langle c_{\alpha,\bk}^\dagger c_{\beta,\bk} \rangle$, which satisfies $P^2 = P$ for Slater determinant states. For GFM states, $P(\bk)$ is $\bk$-independent, and its trace is related to the filling via $\tr P = 4 - |\nu|$. Its spin polarization is $\langle S_i \rangle = \tr s_i P$, where $s_{x,y,z}$ are the spin Pauli matrices. Similarly, the valley polarization (which is the same as the Chern number) is given by $C = \tr \tau_z P$, where $\tau_{x,y,z}$ are the valley Pauli matrices. 

We now discuss the GFM ground states at each integer filling in detail.
At $\nu = -3$, we can write the simple spin-valley ferromagnets as
\begin{equation}
    P^{(3)}_\pm = \frac{1}{4}(1 + s_z)(1 \pm \tau_z).
\end{equation}
This describes a state where valley $\tau = \mp$ is fully empty, whereas valley $\tau = \pm$ has a half-filled spin-polarized with the $\uparrow$ spin filled (see Fig. \ref{fig:strong_coupling_diagrams}). The two states $P^{(3)}_\pm$ are related by time-reversal symmetry and have Chern numbers $C_\pm = \tr P^{(3)}_\pm = \pm 1$. Acting with $\U(2) \times \U(2)$ symmetry yields a manifold of states in each Chern sector that is equivalent to a sphere labelled  by a single unit vector $\bn$. The corresponding density matrix is $P^{(3)}_{\pm,\bn} = \frac{1}{4}(1 + \bn \cdot \bs)(1 \pm \tau_z)$. 

The analysis at $\nu = -1$ is quite similar, with 
\begin{equation}
    P^{(1)}_\pm = \frac{1}{4}(1 + \bn \cdot \bs)(1 \pm \tau_z) + \frac{1}{2}(1 \mp \tau_z)
\end{equation}
which corresponds to completely filling the $\tau = \mp$ valley and half-filling the $\tau = \pm$ valley with a spin-polarized state. This also describes two manifolds of states with Chern number $C_\pm = \mp 1$.

At half-filling, $\nu = -2$, we can write the valley ferromagnets
\begin{equation}
    P_{\tau, \pm}^{(2)} = \frac{1}{2}(1 \pm \tau_z)
\end{equation}
which correspond to filling $\tau = \pm$ valley leading to a QAH insulator with Chern number $C = \pm 2$. Each state is a singlet under $\U(2) \times \U(2)$. In addition, we can also write the spin ferromagnet
\begin{equation}
    P_s^{(2)} = \frac{1}{2}(1 + s_z)
\end{equation}
which has zero Chern number. Acting with $\U(2) \times \U(2)$ generates a manifold of states parametrized by two unit vectors $\bn_\pm$:
\begin{equation}
    P_{s,\bn_+, \bn_-}^{(2)} = \frac{1}{4}(1 + \bn_+ \cdot \bs)(1 + \tau_z) + \frac{1}{4}(1 + \bn_- \cdot \bs)(1 - \tau_z)
\end{equation}
All states $P^{(2)}_{\tau,\pm}$ and $P_{s,\bn_+, \bn_-}^{(2)}$ are degenerate in our model. This degeneracy can be lifted by adding an intervalley Hund's coupling term \cite{YahuiFlatChern, bultinckGroundStateHidden2020} $J_H {\vec S}_+ \cdot {\vec S}_-$ where $\bS_\pm$ are the (second quantized) spin operators in the $\tau = \pm$ valley. For the valley-polarized states $P^{(2)}_{\tau,\pm}$, this term vanishes. For spin polarized states, it yields the anisotropy $ J_H \bn_+ \cdot \bn_-$, which select a ferromagnet $\bn_+ = \bn_-$ for $J_H < 0$ and a spin-valley locked antiferromagnet $\bn_+ = -\bn_-$ for $J_H > 0$.

\begin{table}[h]
    \centering
    {\bf (a) Bandgap [meV]}\\
    \begin{tabular}{cccc}
    \hline \hline
        & $\nu = -1$ & $\nu = -2$ & $\nu = -3$ \\
        \hline
        Strained graphene & 30.97 & 30.64 & 28.82 \\
        Chiral TBG & 22.82 & 17.43 & 12.03 \\
        Realistic TBG & 15.8 & 6.73 & -3.32 \\
        \hline \hline
    \end{tabular}
    \vspace{0.4cm}\\
    {\bf (b) Bandwidth (conduction band) [meV]}\\
    \begin{tabular}{cccc}
    \hline \hline
        & $\nu = -1$ & $\nu = -2$ & $\nu = -3$ \\
        \hline
        Strained graphene & 1.49 & 0.33 & 2.14 \\
        Chiral TBG & 0.25 & 5.64 & 11.04 \\
        Realistic TBG & 1.3 & 9.08 & 19.14 \\
        \hline \hline
    \end{tabular}
    \vspace{0.4cm}\\
    {\bf (c) Bandwidth (valence band) [meV]}\\
    \begin{tabular}{cccc}
    \hline \hline
        & $\nu = -1$ & $\nu = -2$ & $\nu = -3$ \\
        \hline
        Strained graphene & 5.12 & 6.94 & 8.75 \\
        Chiral TBG & 10.54 & 15.94 & 21.33 \\
        Realistic TBG & 21.35 & 31.38 & 41.42 \\
        \hline \hline
    \end{tabular}
    \caption{(a) bandgap, (b) bandwidth for the conduction band, and (c) bandwidth for the valence band for the quasiparticle dispersion for the strong coupling ground states at different filling for our model of strained graphene together with hBN-aligned TBG in the chiral limit $\kappa = 0$ and away from it $\kappa = 0.7$. We use TBG at the first magic angle.}
    \label{tab:EnergyComparison}
\end{table}

As a reference of comparison, we can compare our system to TBG aligned with hBN substrate (TBG-hBN). The TBG-hBN system lacks $C_2$ symmetry and has a single sublattice polarized Chern band per flavor making a direct comparison with our system possible. QAH states have already been observed in TBG-hBN at $\nu=3$ \cite{sharpeEmergentFerromagnetismThreequarters2019a, serlinIntrinsicQuantizedAnomalous2020a}, and FCIs were observed at finite but small magnetic field \cite{xieFractionalChernInsulators2021a}. To make our comparison transparent, we use same energy units for both systems, i.e. we use energy units $E_0 = 0.324$ eV, which corresponds to $L_M \approx 13.3$ nm, the moir\'e period at TBG's first magic angle. Fig.~\ref{fig:SpectrumComparison} and Table~\ref{tab:EnergyComparison} contain a comparison between our system and TBG-hBN both in the ideal chiral limit and the more realistic limit where the ratio of intrasublattice to intersublattice tunneling is finite, $\kappa = 0.7$. We conclude that our system has flatter Chern bands than TBG, even in the presence of interactions. We therefore expect it to host a variety of strongly-correlated phases.

\end{document}